\documentclass[%
 twocolumn,
 amsmath,amssymb,
 aps,
prb,
]{revtex4-2}

\usepackage{soul, color} 
\usepackage{graphicx}
\usepackage{dcolumn}
\usepackage{bm}

\usepackage[normalem]{ulem}
\usepackage{xcolor}
\begin{document}

\preprint{APS/123-QED}

\title{Observation of an accidental bound state in the continuum \\ in a chain of dielectric disks}

\author{M.\,S. Sidorenko$^{1}$}
\email{mikhail.sidorenko@metalab.ifmo.ru}
\author{O.\,N. Sergaeva$^1$}
\author{Z.\,F. Sadrieva$^1$}
\author{C. Roques-Carmes$^2$}
\author{P.\,S. Muraev$^{3,4}$}
\author{D.\,N. Maksimov$^{3,4}$}
\author{A.\,A. Bogdanov$^{1,}$}
\email{a.bogdanov@metalab.ifmo.ru}
\affiliation{  
$^1$Department of Physics and Engineering, ITMO University, 197101 St. Petersburg, Russia  \\
$^2$Research Laboratory of Electronics, Massachusetts Institute of Technology, 50 Vassar St., Cambridge, MA \\
$^3$Kirensky Institute of Physics, Federal Research Center KSC SB RAS, 660036 Krasnoyarsk, Russia \\
$^4$Siberian Federal University, 660041 Krasnoyarsk, Russia \everypar{\looseness=-1}
}

\date{\today}

\begin{abstract}

Being a general wave phenomenon, bound states in the continuum (BICs) appear in acoustic, hydrodynamic, and photonic systems of various dimensionalities. Here, we report the first experimental observation of an accidental electromagnetic BIC in a one-dimensional periodic chain of coaxial ceramic disks. We show that the accidental BIC manifests itself as a narrow peak in the transmission spectra of the chain placed between two loop antennas. We demonstrate a linear growth of the radiative quality factor of the BICs with the number of disks that is well-described with a tight-binding model. 
We estimate the number of the disks when the radiation losses become negligible in comparison to material absorption and, therefore, the chain can be considered practically as infinite. The presented analysis is supported by near-field measurements of the BIC profile. The obtained results provide useful guidelines for practical implementations of structures with BICs opening new horizons for the development of radio-frequency and optical metadevices.
\end{abstract}

\maketitle

\newcommand{\ph}{\varphi}

\section{Introduction}
Dielectric resonators are open systems whose eigenmodes couple to the radiation continuum. Due to the coupling, the eigenmodes are generally expected to have finite lifetimes because of radiation losses. For a long time, it was believed that only guided modes
with frequencies below the light line were decoupled from the radiation continuum~\cite{phc,Fan1999}.
Nevertheless, in the early 2000's, several counterexamples of perfectly localized states -- i.e. totally decoupled from the radiation continuum -- at frequencies above the light line were proposed in dielectric gratings and photonic crystal waveguides~\cite{paddon2000two,bulgakov2008bound,marinica2008bound}.
Historically, a similar phenomenon was first predicted for quantum particles trapped above quantum well barriers~\cite{Neumann}. Such localized states are known as \textit{bound states in the continuum (BIC)}. 

A genuine BIC is a mathematical idealization, that displays an infinite quality factor ($Q$-factor), which can only be observed in infinite periodic  systems or in finite systems with exploiting epsilon-near-zero materials~\cite{monticone2014embedded}. In practice, one can only observe quasi-BICs with finite $Q$-factor. The $Q$-factor is limited due to several factors such as the finite size of the resonator~\cite{sadrieva2019experimental,taghizadeh2017quasi,Blaustein:07,Bulgakov2017}, structure imperfections~\cite{Ni:2017,jin2019topologically}, influence of the substrate~\cite{hsu2013observation,sadrieva2017transition}, surface roughness~\cite{sadrieva2017transition}, symmetry breaking~\cite{koshelev2018asymmetric,hsu2013observation} and material losses~\cite{sadrieva2017transition,sadrieva2019experimental,PhysRevA_Hu2020}. Recent progress in photonic crystals fabrication and characterization stimulated extensive studies of BICs in various periodic photonic structures with potential applications to novel lasing sources~\cite{Kodigala2017,ha2018directional}, sensing~\cite{romano2017optical,liu2017optical,romano2018surface,Koshelev2019}, optical filtering~\cite{foley2014symmetry,cui2016normal,doskolovich2019integrated}, enhanced light-matter interaction ~\cite{koshelev2018strong, kravtsov2020nonlinear, yoon2015critical,mocella2015giant,zhang2015ultrasensitive, magnusson1992new}, nonlinear nanooptics~\cite{wang2018large,Koshelev2019,wang2017improved}, spintronics~\cite{ramos2014bound}, and optical vortex generation~\cite{doeleman2018experimental,wang2020generating}.

Among the variety of available periodic photonic systems, a one-dimensional array of spheres or disks stands out because of its rotational symmetry. This gives rise to BICs \textcolor{black}{with non-trivial orbital angular momentum (OAM)~\cite{bulgakov2014bloch,bulgakov2016light}}. Such states could be used to generate vortex beams~\cite{doeleman2018experimental,Wang2020} with potential applications in optomechanics~\cite{padgett2011tweezers}, and quantum cryptography. The theory of BIC with OAM was developed in Refs.~\cite{bulgakov2015light,bulgakov2017bound,Bulgakov_OAM2017}. The first experimental observation of a symmetry-protected BIC with zero OAM was reported in our previous work~\cite{sadrieva2019experimental}. So far all experimentally-reported BICs in one-dimensional arrays fell into the category of in-$\Gamma$ BICs. Such BICs are standing waves that do not couple with outgoing radiative waves due to symmetry. 

In this work, we focus on off-$\Gamma$ or accidental \cite{hsu2013observation} BICs which
are traveling Bloch waves propagating along the ceramic disks chain. While the in-$\Gamma$ BICs are insensitive to variation of the system's parameters, the experimental detection of an accidental BIC is more challenging, since it requires fine adjustment of the system's parameters: radius, height of disks, period, or permittivity. Here, we analyze the transformation of a resonant state into an accidental off-$\Gamma$ BIC with zero OAM in an axially symmetric 1D periodic  array of dielectric disk resonators. We experimentally demonstrate this transformation in the GHz-frequency range. 

The paper is organized as follows: in Sec.~\ref{sec:inf} we discuss the eigenmode spectrum of the infinite chain with both in-$\Gamma$ and off-$\Gamma$ BICs. In Sec.~\ref{sec:finite} we turn to the realistic chain of a finite length and discuss the transformation of high-$Q$ resonances to BIC with the increase of the disk numbers. We also introduce a tight-binding model that well describes this transformation.  Section~\ref{sec:exper} contains the experimental results confirming the theoretical predictions.

\begin{figure*}[t]
    \centering
    \includegraphics[width =0.85\linewidth ]{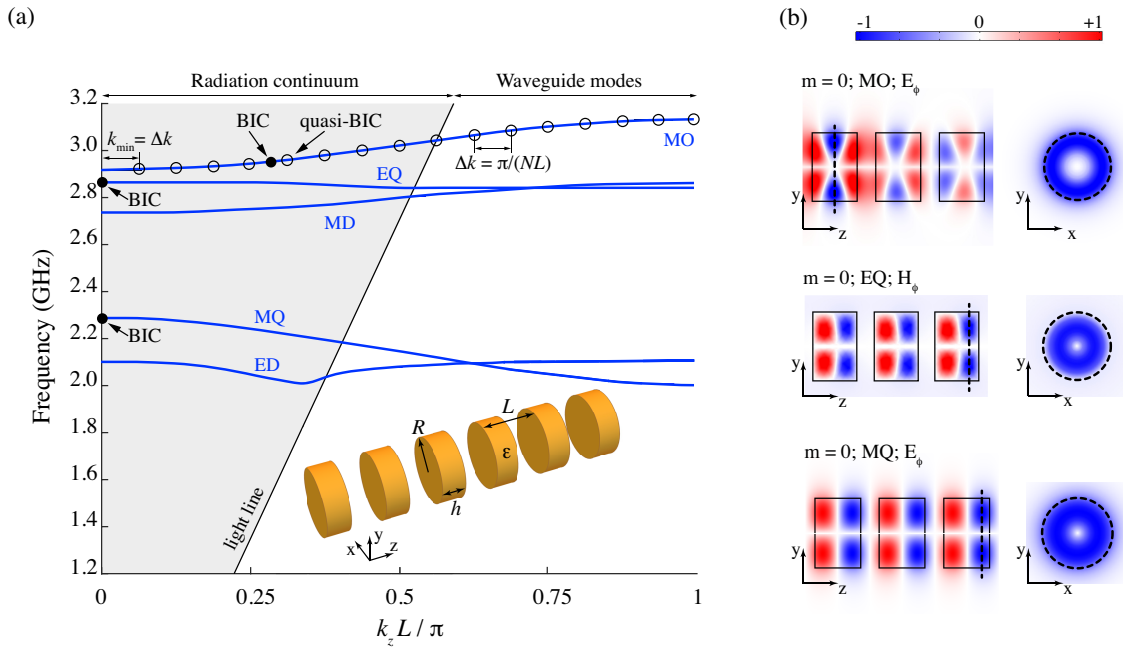}
    \caption{Bound state in the continuum in the chain of ceramic disks with period $L=28$~mm, radius $R=15$~mm, height $h=20$~mm and permittivity $\varepsilon=44$ -- (a) Photonic band structure of an infinite chain of disks with symmetry-protected and accidental BICs. Several modes with zero OAM $m=0$ are shown: $ED (MD)$ -- electric (magnetic) dipole moment, $EQ (MQ)$ -- electric (magnetic) quadrupole moment and $MO$ -- magnetic octupole moment. The inset shows a 3D rendering of the chain. (b) Side and front view of the field distribution for the BICs.} 
    \label{fig:1}
\end{figure*}

\section{Infinite chain -- Theory}\label{sec:inf}

First, we consider an infinite periodic chain of ceramic disks shown in Fig.~\ref{fig:1}(a). Applying Bloch's theorem and taking into account the axial rotational symmetry of the system, the electric field of an eigenmode can be written as
\begin{equation}
    \textbf{E}(r,\ph,z,t) = \textbf{U}_{m,k_z}(r, z)  ~e^{-i\omega t\pm ik_z z\pm im\ph},
\end{equation}
where $k_z$ is the Bloch wave vector defined in the first Brillouin zone, $m$ (an integer number) is the OAM, and $\mathbf{U}_{m,k_z}$ is the periodic Bloch amplitude. Due to the reflection symmetry $z\rightarrow -z$ the eigenmodes propagating in $+z$ and $-z$ directions are degenerate, which results in $\pm$ sign in the argument of the exponential. Since the function $\textbf{U}_{m,k_z}(z,r)$ is periodic in $z$ with period $L$, it can be expanded into a Fourier series:
\begin{subequations}
\begin{equation}
    \textbf{U}_{m,k_z}(r,z) = \sum_n \textbf{C}^n_{m,k_z}(r)  ~e^{\frac{2\pi i n}{L}z},
\end{equation}
where
\begin{equation}
    \textbf{C}^n_{m,k_z}(r) = \frac{1}{L} \int\limits_{-\frac{L}{2}}^{\frac{L}{2}} \textbf{U}_{m,k_z}(r,z) ~e^{-\frac{2\pi i n}{L}z} dz.
\end{equation}
\end{subequations}
The index $n$ is an integer and it corresponds to the diffraction order. Generally, the Fourier coefficient $\textbf{C}^n_{m,k_z}(r)$ defines the amplitude of the near-field or outgoing wave carrying energy away from the structure. In the subwavelength regime $L <\lambda$, only the zero-th diffraction order ($n=0$) is non-zero. 

Figure~\ref{fig:1}(a) shows the dispersion of the modes with $m=0$. The simulation is performed for the chain with period $L=28$~mm composed of disks with diameter $D=30$~mm, height $h=20$~mm and permittivity $\varepsilon = 44$. The grey area depicts the domain where only the zero-th diffraction order is open. Thus, if $k_z <\omega/c<|k_z\pm2\pi/L|$ the radiation losses are determined by the zero-order Fourier coefficient only. If the zero-order Fourier amplitude $\textbf{C}^0_{m,k_z}(r)$ equates to zero, the leaky resonant eigenmode turns into a BIC~\cite{hsu2013observation}. By  definition, the zero-order Fourier coefficient is the spatial average of the electric field over a period. The center of the Brillouin zone is the $\Gamma$-point, where the symmetry of the field distribution follows the point symmetry of the chain, i.e. modes transform by the irreducible representations of the chain's point-group symmetry~\cite{Ivchenko1995, Sakoda}. Since the chain is invariant under reflection in the plane $z=0$, the modes are divided into odd and even with respect to the $z\rightarrow -z$ transformation. Obviously, the integration of an odd function $\textbf{U}_{m,k_z}(r,z)$ with respect to $z$ yields zero coupling. In this case, we observe a symmetry-protected BIC with the coupling to the radiation continuum prevented by the point symmetry of the chain. 

\begin{figure*}[t]
    \centering
    \includegraphics[width =1\linewidth ]{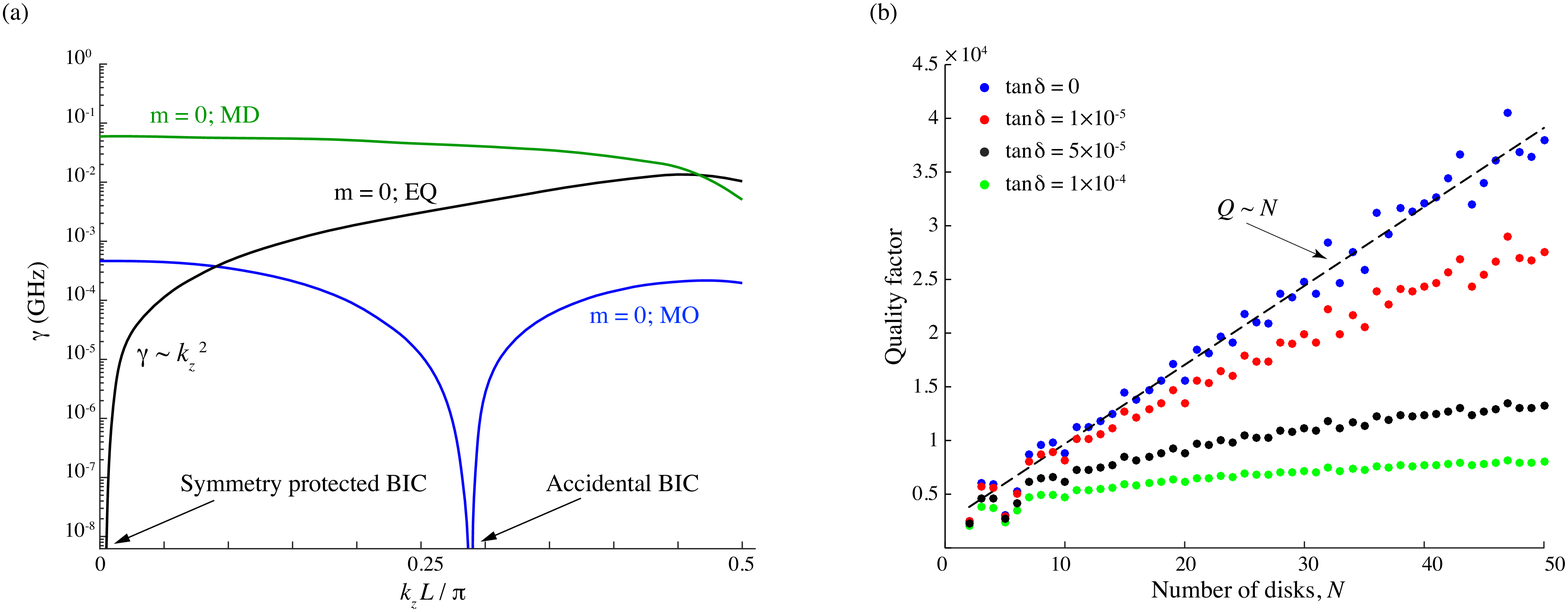}
    \caption{(a) Radiation losses of three selected modes in the infinite chain, including the symmetry-protected and off-$\Gamma$ accidental BICs. (b) Dependence of the $Q$-factor of the accidental quasi-BIC as a function of the number of disks for different levels of material losses}. 
    \label{fig:2}
\end{figure*}

In Fig.~\ref{fig:1}(a), we show that the odd modes are formed by the electric ($EQ$) and magnetic ($MQ$) quadrupole Mie resonances whose directivity patterns are the $xy$-nodal surface. Therefore, these modes do not contribute to the far-field in the direction of a single open diffraction channel coinciding with the $y$-axis. In contrast, the even modes such as the  magnetic octupole ($MO$), magnetic ($MD$), and electric ($ED$) dipoles radiate in the $y$-direction. The dispersion of the radiation losses of $MD$, $EQ$ and $MO$ modes characterized by $\gamma = \text{Im}[\omega(k_z)]$ is shown in Fig.~\ref{fig:2}(a). While $\gamma$ remains constant for the $MD$ mode, the radiation loss of the $EQ$ mode limits to zero as $k_z^2$ in the vicinity of the $\Gamma$-point revealing the emergence of a symmetry-protected BIC.\\

The spatial average of non-even functions over the unit cell may vanish at an arbitrary point $k_z$ \textit{accidentally} resulting in appearance off-$\Gamma$ BIC also called {\it accidental BIC}). For instance, the $MO$ mode turns to the accidental BIC at the specific $k_z$ whose field distribution is shown in  Fig.~\ref{fig:1}(b). The leakage vanishes when all multipoles forming the mode interfere destructively in the direction of the open diffraction channel~\cite{sadrieva2019multipolar}. The accidental BIC on the $MO$ dispersion branch manifests itself in the vanishing of the radiative losses as shown in Fig.~\ref{fig:2}(a).

It is known that the states with nonzero OAM ($m\neq0$) exhibit hybrid polarization, while modes with $m=0$ can be divided into transverse electric (TE) and transverse magnetic (TM)~\cite{snyder2012optical,AdvEM}. Hereinafter, TE (resp. TM) refers to a field decomposition of the form $\textbf{E} = (0,E_\ph,0)$ and $\textbf{H} = (H_r,0,H_z)$ (resp. $\textbf{E} = (E_r,0,E_z)$ and $\textbf{H} = (0,H_\ph,0)$). For a hybrid polarization $C_{\text{TE}}\cdot(H_r,E_\ph,H_z) + C_{\text{TM}}\cdot(H_r,H_\ph,E_z)$, an additional degree of freedom $C_{\text{TE(TM)}}$ arises in the system. 

As long as for the particular polarization the parity of $E_{(\cdot)}$ and $H_{(\cdot)}$ components are opposite, it is impossible to obtain a zero spatial average of the electric and magnetic field components simultaneously~\cite{AdvEM}. Therefore, a symmetry-protected BIC cannot be observed in a band with nonzero OAM. However, an accidental BIC can be obtained by tuning geometrical and material parameters of the structure, as observed in Fig.~\ref{fig:1}. 

\section{Finite chain -- Theory}\label{sec:finite}
\subsection{Fabry-Perot quantization}
 In experiments, we always deal with structures of finite sizes. In this case, the ends of the chain play the role of partially-reflecting mirrors and, thus, the chain can be considered as a Fabry-Perot resonator. The continued frequency band of the infinite chain is quantized into Fabry-Perot resonances with finite $Q$-factors limited by the scattering from the ends of the chain. Therefore, in a finite chain, a genuine BIC turns into a quasi-BIC with a finite $Q$-factor.
 
For the chain of $N$ scatterers, the continuous band of the infinite chain is replaced by a finite set of $N$ Fabry-Perot resonances at frequencies $\omega(k_z)$ corresponding to the quantized quasi-wave vector $k_z = p \pi/(NL)$, where $p$ is an integer, and $L$ is the period. The highest $Q$-factor resonant state has the quantized wave vector closest to the BIC (observed in the infinite periodic system) and, thus, can be associated with the quasi-BIC [see Fig.~\ref{fig:1}(a)]. Similarly, a symmetry-protected quasi-BIC coincides with the first resonant state related to the minimal wave vector $k_z = \pi/(NL)$ as it was shown previously in our paper~\cite{sadrieva2019experimental}. In the case of an off-$\Gamma$ BIC, though, the index $p$ of the nearest to BIC resonance is changing with the increase of the number of the disks. The change of $p$ also explains the fluctuation of the $Q$-factor around the mean; the distance in the $k$-space between the genuine BIC and the quasi-BIC can take arbitrary values in the interval $[-\pi/(2NL), \pi/(2NL)]$.

Our simulation results for the chain with a finite length, the $Q$-factor of the resonant state associated with the accidental out of $\Gamma$ quasi-BIC can be approximated by a linear function of a number of disks $N$, see Fig.~\ref{fig:2}(a). The experimental study of this dependence is provided in Sec.~\ref{sec:exper}. Here, we highlight that the $Q$-factor of a symmetry-protected quasi-BIC grows as $N^2$~\cite{Bulgakov2017,sadrieva2019experimental}. In contrast, the $Q$-factor of an accidental quasi-BIC at the $\Gamma$-point scales as $N^3$ with the number of disks~\cite{Bulgakov2017,Bulgakov2019}. To explain the difference in the asymptotic behaviour of the $Q$-factor we introduce a simple tight-binding model that accounts for partial reflections from the ends of the chain. Using the tight-binding model is quite natural as the disks are made of high-refractive-index ceramics disks tightly trapping the Mie modes inside [see Fig.~\ref{fig:1}(b)].

\subsection{Tight-binding model}


In a finite chain, the propagating Bloch solutions reflect from the ends of the chain, forming Fabry-Perot resonances -- standing waves with a quantized propagation constant $k_z$. As we are interested only in the behavior in the vicinity of the accidental BIC, we only account for the radiation losses at the ends of the chain neglecting losses due to detuning from the BIC point. 
Mathematically, the system described above can be formulated as the following eigenvalue problem
\begin{equation}\label{Apa1}
  (\widehat{H}_{{\rm eff}}-\omega)|\psi\rangle = 0,
\end{equation}
where $\widehat{H}_{{\rm eff}}$ is a non-Hermitian operator describing both wave propagation between the mirrors and radiation losses when reflecting from the edges of the array. Equation (\ref{Apa1}) can be introduced by projecting the full-wave Maxwell's equation onto the Wannier states of the guided modes. Thus, the amplitude of the electromagnetic field at the $q_{\rm th}$ site is given by 
\begin{equation}
\psi (q)=\langle q| \psi \rangle,
\end{equation}
where $|q\rangle$ is the Wannier function localized at this site. 
The operator $\widehat{H}_{{\rm eff}}$ can be defined by its action on the array of the local field amplitudes
\begin{equation}\label{Apa2}
-\frac{1}{2}\left(\begin{array}{cccc}
i\eta & J & \ldots & 0 \\
J & 0 & \ldots & 0 \\
\vdots& \vdots &\ddots & \vdots\\
0 & 0 & \ldots & i\eta
\end{array}\right)
\left( \begin{array}{c}
\psi (1) \\
\psi (2) \\
\vdots \\
\psi (N)
\end{array}\right) =
\omega
\left( \begin{array}{c}
\psi (1) \\
\psi (2) \\
\vdots \\
\psi (N)
\end{array}\right),
\end{equation}
where $J$ describes the optical coupling between the sites (discs) and $\eta$ 
accounts for the radiation losses in reflection from the edges. For simplicity, we only introduced nearest-neighbour couplings in Eq. (\ref{Apa2}). Based on our numerical findings we can conclude that $\eta \ll J$.

Let us look for the radiative eigenmode in the form of a superposition of two counter-propagating waves
\begin{equation}\label{Apa3}
\psi(q) = Ae^{ikq} + Be^{-ikq}
\end{equation}
with dispersion
\begin{equation}\label{Apa4}
\omega=-J\cos(k).
\end{equation}
After substituting Eq. (\ref{Apa3}) into Eq. (\ref{Apa2}) one finds that $k$ satisfies the following equation
\begin{equation}\label{Apa8}
\sin[k(N+1)]-2i\frac{\eta}{J}\sin[kN]-\left(\frac{\eta}{J}\right)^2\sin[k(N-1)]=0.
\end{equation}
Let us introduce a series expansion of $k$ in the powers of $\eta/J$
\begin{equation}\label{Apa9}
k = k^{(0)}_p (N) + \alpha_p(N) \frac{\eta}{J} + {\cal O}\left[\left(\frac{\eta}{J}\right)^2\right],
\end{equation}
where
\begin{equation}\label{Apa10}
k^{(0)}_p (N)=\frac{\pi p}{N+1}, \ p=1, 2, \ldots, N
\end{equation}
is the zero-th order solution corresponding to the lossless chain, and
$\alpha_p(N)$ gives the correction to the wavevector
in the first perturbation order. 
After substituting Eq. (\ref{Apa9}) into Eq. (\ref{Apa8})
and collecting the terms of the same order in $\eta/J$ one
finds
\begin{equation}\label{Apa11}
\alpha_p(N)=2i(-1)^{p}\frac{\sin[k^{(0)}_p(N)]}{N+1}.
\end{equation}
Notice that $\alpha_p(N)$ is imaginary since it is solely due to the radiative losses.
Finally, the inverse lifetime $\gamma_p=\text{Im}\{\omega_p\}$ is found by substituting Eq. (\ref{Apa9}) into Eq. (\ref{Apa4})
\begin{equation}\label{Gamma1}
\gamma_p=\frac{\eta}{N+1}\sin^2\left[k^{(0)}_p(N) \right].
\end{equation}
One can evaluate the $Q$-factor scaling law against $N$ by examining the asymptotics of Eq.~(\ref{Gamma1}). For an in-$\Gamma$ resonant state, we take $p=1$, and $k^{(0)}_1(N)=\pi/(NL)$. Thus, it follows from Eq.~(\ref{Gamma1}) that $\gamma_1\propto 1/N^{3}$ which is in accordance with the previous findings by Blaustein {\it et al} \cite{Blaustein:07}. In the case of an off-$\Gamma$ resonant state, however, we have to keep $k^{(0)}_p(N)$ fixed. Then, Eq.~(\ref{Gamma1}) yields 
\begin{equation}
\gamma\propto 1/N.
\end{equation}
Note that we do not use subscript in the above equation as the resonances with the wave number nearest to given $k_z$ always have a different number $p$. 

As it was mentioned above our model neglects the radiation losses due to $k_z$ detuning. Generally, the total losses of a Fabry-Perot resonance in the chain are comprised of radiation at the ends of the chain and of side radiation existing due to the leaky nature of the propagation band. The total radiative $Q$-factor, $Q_{{\rm rad}}$ is then given by
\begin{equation}\label{Qrad0}
    {Q_{{\rm rad}}}=\frac{Q_1Q_2}{Q_1+Q_2},
\end{equation}
where $Q_1$ is due to the radiation at the edges, while $Q_2$ accounts for radiation by the leaky band itself. Notice that $Q_2$ is also a diverging quantity as the BIC point is separated from the nearest resonance by a distance of less than $\pi/(2NL)$ as seen from Fig. \ref{fig:1}(a). Then, for $Q_2$ we have
\begin{equation}
    Q_2\propto N^2,
\end{equation}
since the dispersion of the radiation rate must be quadratic in the spectral vicinity of a BIC because the radiation rate is non-negative. Therefore, with $N\rightarrow\infty$ we have
\begin{equation}\label{Qrad}
    {Q_{{\rm rad}}}\propto N.
\end{equation}
We mention in passing that the dispersion of the radiation rate in the vicinity of an in-$\Gamma$ accidental BIC is quartic \cite{yuan2017strong, bulgakov2017topological}. Then, according to Eq.~(\ref{Qrad0}) the edge losses dominate in the asymptotic behaviour and we have ${Q_{{\rm rad}}}\propto N^3$ as shown in \cite{Bulgakov2017}.

\subsection{Material losses}

Besides the radiation losses due to the finite size of the chain, there are other sources of losses (absorption in material, roughness of the scatterers, structural disorder, leakage into the substrate, etc.) which contribute to the total losses even for an infinite chain~\cite{Ni:2017,sadrieva2017transition,PhysRevA_Hu2020}. In our case, the loss mechanism limiting the total $Q$-factor at a large number of disks ($N\gg1$) is absorption in the ceramics. Therefore, the total $Q$-factor ($Q_{\text{tot}}$) can be found as
\begin{equation}\label{Qtot}
Q_{\text{tot}}^{-1}=Q_{\text{rad}}^{-1}+Q_{\text{abs}}^{-1},    
\end{equation}
where $Q_{\text{abs}}\approx1/\tan\delta=\text{Re}(\varepsilon)/\text{Im}(\varepsilon)$.

Figure~\ref{fig:2}(b) shows the dependence of  $Q_\text{tot}$ factor of the quasi-BIC on the number of disks $N$ for various $\tan\delta$ -- tangent of material losses. For a large number of disks, $Q_\text{tot}$ saturates, i.e. the material losses give the major contribution to the total losses. Practically, we can determine how many disks in the chain are necessary to neglect finite-size effects with respect to other sources of losses. This question is discussed in the next section.

\begin{figure}[t]
    \centering
    \includegraphics[width =1\linewidth ]{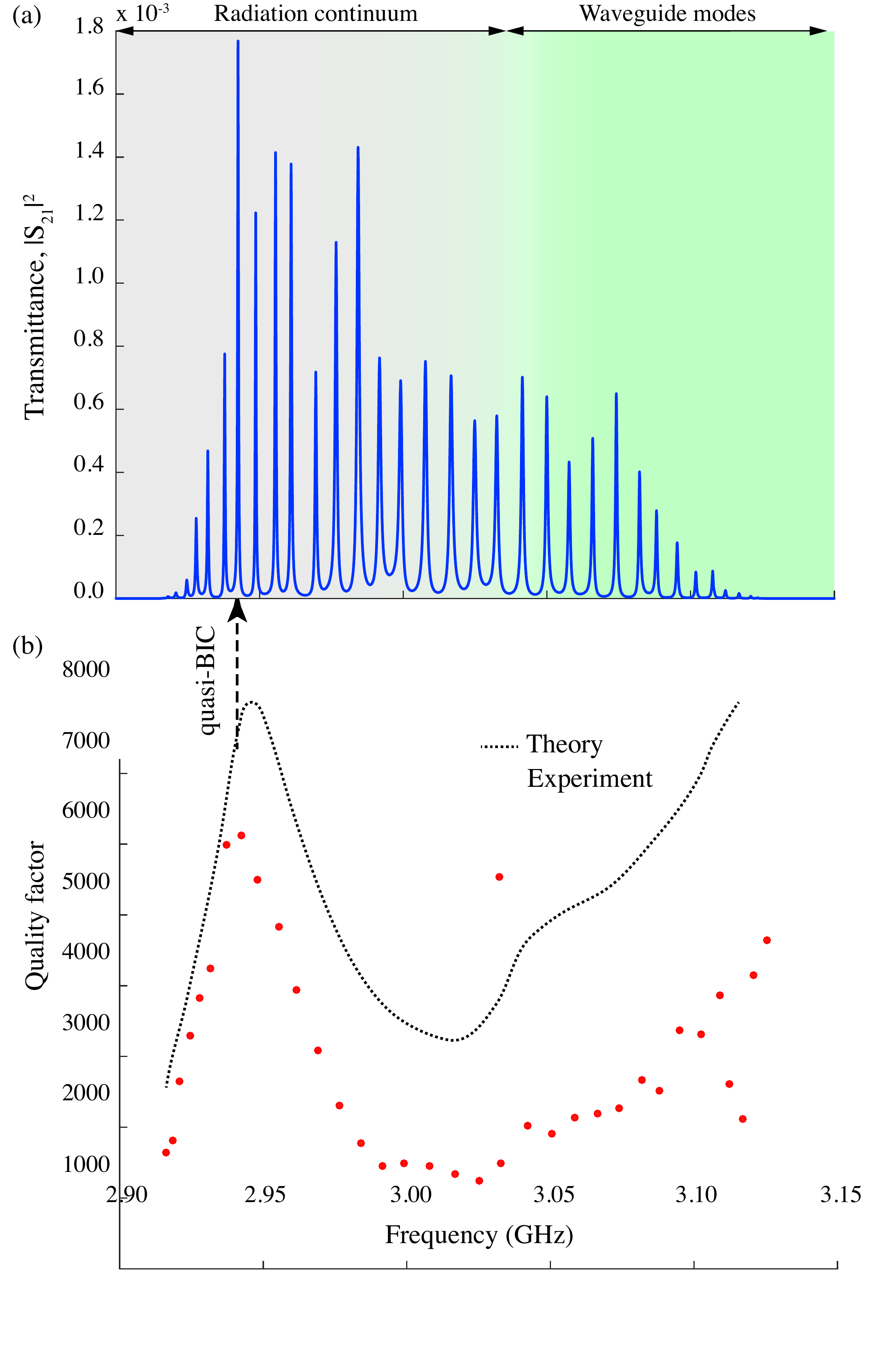}
    \caption{(a) Transmission spectrum of the 36-disk chain. The magnetic octupole (TE) mode with $m=0$ was excited with an electric dipole antenna. The green color shaded area represents the frequency domain of waveguide modes, while the grey shaded area the radiation continuum. (b) The $Q$-factor of the resonant states of the chain composed of the 36 disks. The accidental quasi-BIC is observed at the frequency about 2.95~GHz. }
    \label{fig:3}
\end{figure}

\section{Experimental demonstration}\label{sec:exper}
\subsection{Transmission measurement}
\begin{figure}
    \centering
    \includegraphics[width =1\linewidth ]{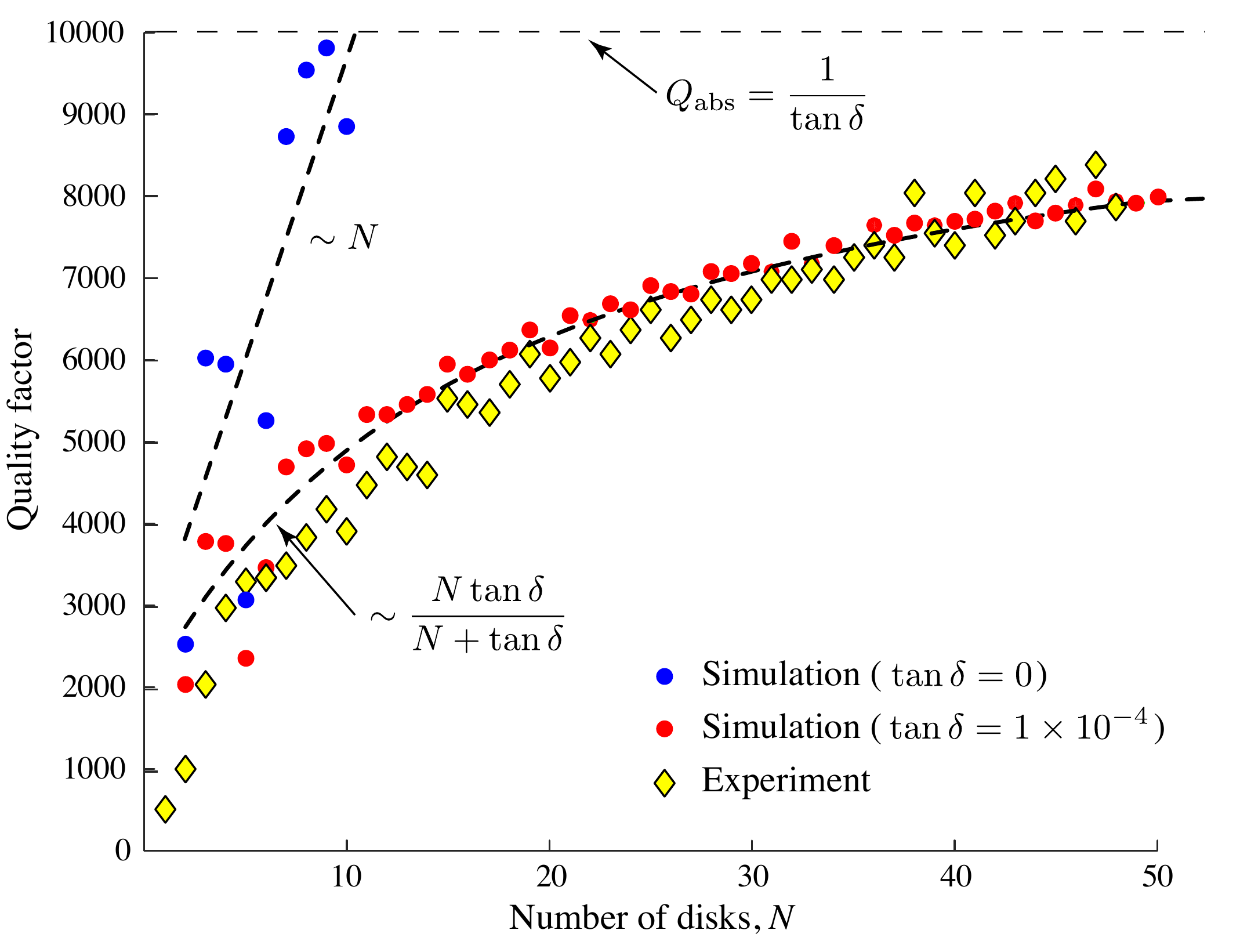}
    \caption{Experimental and theoretical dependence of the $Q$-factor of the accidental quasi-BIC. The dashed line corresponds to the predicted scaling lax for the lossless disk chain.} 
    \label{fig:4}
\end{figure}
\begin{figure}
    \centering
    \includegraphics[width =0.7\linewidth ]{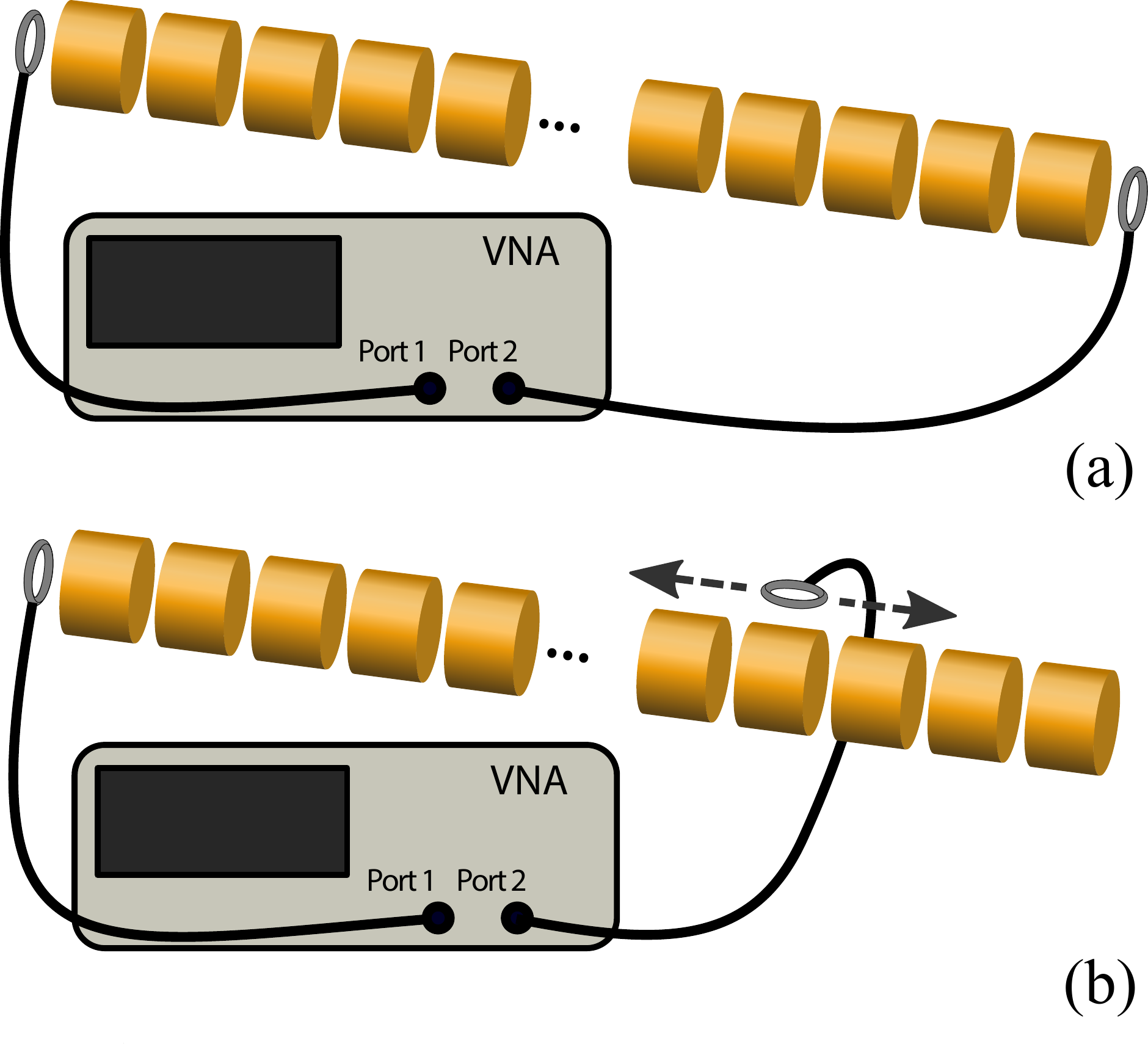}
    \caption{(a) Experimental setup for transmission spectrum measurement. (b) Experimental setup for measurement of the radial component of the magnetic field.} 
    \label{fig:5ab}
\end{figure}

A chain of maximum 48 disks with the period $L=28$~mm, where each disk is made of a ceramic material with $\varepsilon=44$ and $\tan \delta=1\cdot10^{-4}$ with dimensions $D=30$~mm, $h=20$~mm was considered. (For more details, see Appendix). As it was mentioned in Sec.~\ref{sec:inf}, the two polarization components of a mode with $m=0$ can be distinguished completely. Therefore, TE and TM modes can be excited separately by magnetic or electric dipole antennas, respectively. 
We excite the magnetic octupole (TE) mode via near-field coupling with a magnetic dipole antenna. Two shielded loop antennas are placed along the same axis in 5~mm distances from the corresponding faces of the first and the last cylinders of the chain. The resonant frequencies of each antenna are higher than the frequency band of our interest. Such antenna placement allows us to assume weak-coupling, which makes the analysis of the $Q$-factor eligible. The scheme of the transmission spectrum measurement setup is provided in Fig.~\ref{fig:5ab}(a). Both antennas are connected to the corresponding ports of a Vector Network Analyser (VNA) Rohde$\&$Schwarz. The transmission spectrum is measured as the $S_{21}$ parameter from the VNA. The measured frequency band is from 2.9 GHz to 3.15~GHz, with $3.2\cdot10^4$ frequency samples taken in this interval. This large number of frequency samples ensures sufficient resolution to carefully observe high-$Q$ modes in the transmission spectrum. 

A sample of a measured transmission spectrum for the chain of 36 cylinders is plotted in Fig.~\ref{fig:3}(a). The resonances lying in the green area correspond to the waveguide modes, those from the grey area are leaky modes coupled to the radiation continuum. 

The loaded $Q$-factor is then extracted as $Q_{\text{loaded}}=2\Delta f/f_0$, where $\Delta f$ is the half-width of the transmission spectrum local maximum at frequency $f_0$. The width is measured at the level 0.7 of the local maximum value. The measured values of the loaded $Q$-factor for a 36-disk chain are plotted as a function of resonant frequency in Fig.~\ref{fig:3}(b). On the same plot, a theoretical curve for the $Q$-factor values is also presented, obtained from the theory developed in Sections II and III. The sharpest peak in the transmission spectrum corresponds to the accidental quasi-BIC. The sufficiently high intensity of the quasi-BIC indicates that it couples efficiently to the loop antennas placed at the end of the chain. Moreover, the theoretical curve in Fig.~\ref{fig:3}(b) (dot line) shows that structural disorder has a negligible effect. Despite the leakage and absorption, we clearly observe the maximum in $Q$ vs frequency indicating the manifestation of accidental BIC.

 Then, we measure the transmission spectra for the chain consisting of a different number of disks $N$ and extract the value of $Q$-factor corresponding to accidental quasi-BIC.    
The resulting experimental dependence $Q$ vs $N$ is shown in Fig.~\ref{fig:4}. The experimental curve fits the theoretical curve very well, see Eq.~(\ref{Qtot}). Although the theory predicts the linear dependence of the $Q$-factor on the number of disks expressed by Eq.~(\ref{Qrad}), the material losses result in an essential deviation from the $Q\sim N$ law. When increasing the number of disks, the total $Q_\text{tot}$ factor saturates at $Q_{\text{abs}} = 1/\tan\delta$.  With smaller material losses, one could achieve a larger $Q$-factor in the limit of long chains. The analyzed chain of ceramics disks is characterised by $\tan\delta = 1\cdot 10^{-4}$. Therefore, the saturation level $Q_{\text{abs}} = 10^4$ shows the maximal $Q_\text{tot}$ factor that can be obtained for the infinitely long chain.

One can see that from $N$ equals about 30, the $Q$-factor increases with the number of disks slowly. Therefore, we suppose that for the experimental measurements about 30 or more disks is enough to demonstrate the BIC phenomenon.

\begin{figure*}[t]
    \centering
    \includegraphics[width =0.9\linewidth ]{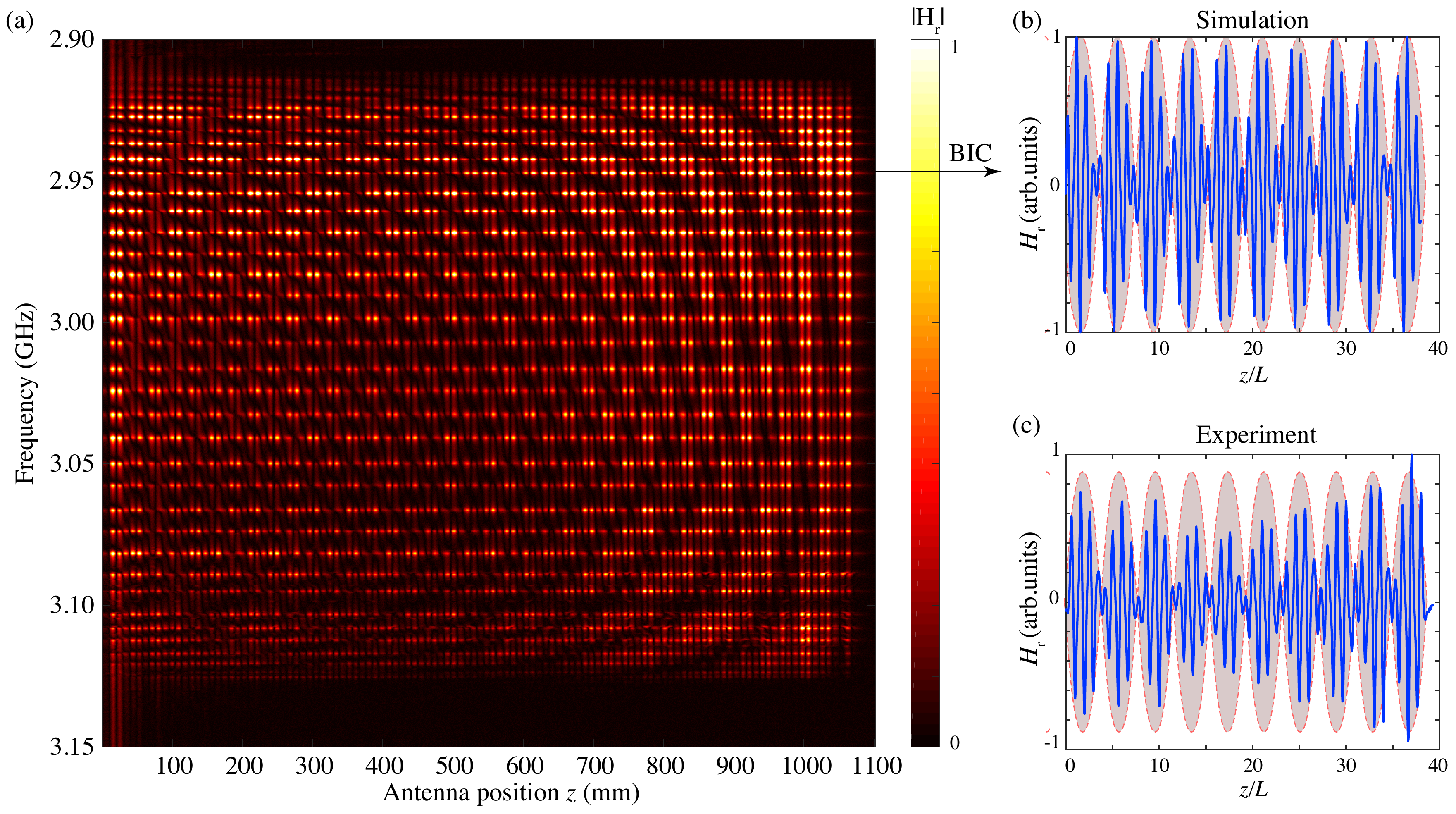}
    \caption{(a) Measured radial component of the magnetic field as a function of the antenna position $z$ and frequency. The antenna was moved along the chain. Panels (b) and (c) show magnetic field distribution obtained in simulation and experiment at the frequency of BIC.} 
    \label{fig:6}
\end{figure*}

\subsection{Field distribution measurements}
We also measured the spatial and spectral distribution of the radial component of the magnetic field along the chain of cylinders measured. The measurements were performed in the near-field by using a magnetic loop antenna with the main axis coinciding with the cylinder radius [see Fig.~\ref{fig:5ab}(b)]. The distance between the antenna and the cylinder wall was 2~mm. The field in the chain was excited via near-field coupling by another loop antenna, located along the same axis, at the end of the disk chain. The receiving antenna was positioned at different positions along the chain by means of a PC-controlled high-precision near-field scanner. Both antennas were connected to the ports of a VNA, which took one transmission spectrum scan in a frequency band between 2.9 GHz and 3.15 GHz for every spatial position of the receiving antenna. In such a setup the modulus of the transmission spectrum is proportional to the modulus of the radial component of the magnetic field in the near-field zone of the sample. The receiving antenna was positioned in steps of 0.5~mm. The normalized spectral and spatial distribution is shown in~Fig.~\ref{fig:6}. The field profile for the selected frequency of 2.94~GHz is presented in the same figure in insets. The results of the numerical simulation for the radial component of the magnetic field distribution are also presented, showing good agreement with our experiment.

\section{Conclusion}
We have observed an accidental BIC in a one-dimensional periodic array composed of ceramics disks. In the experiment, we selectively excite a magnetic octupole mode with zero orbital angular momentum and measured the transmission spectrum using the coaxially placed loop antennas. We extracted the $Q$-factor from the experimental data for the arrays with varying number of disks and revealed a linear growth of the $Q$-factor on the number of disks confirmed by our tight-binding model. For a certain ceramic with a material loss $\tan \delta = 1\cdot 10^{-4}$, we found that radiation losses become negligible in comparison to material absorption when the number of disks is about 30 and, therefore, that the chain can be practically considered as infinite. The obtained results provide useful guidelines for practical implementations of structures with BICs that open up new horizons for the development of radiofrequency and optical metadevices.\\

\section*{Acknowledgement}
The Authors acknowledge Almas Sadreev for fruitful discussions. M.D.N. acknowledges the financial support  by the Government of the Russian Federation through the ITMO Fellowship and Professorship Program and thanks ITMO University for hospitality. C. R.-C. acknowledges funding from the MIT MISTI-Russia Program.

\section*{APPENDIX}
We consider an experimental sample consisting of a finite chain of maximum 48 disks. Each disk is made of microwave ceramics with nominal dielectric permittivity of $\varepsilon=44$ and tangent loss of $\tan \delta=1\cdot 10^{-4}$ in the frequency region 2-4~GHz. Each disk is 30 mm-long and has a diameter of 30 mm. All the cylinders in the chain are aligned along the same axis and equidistant with a period of $L=28$~mm. To ensure that all the disks are placed exactly at a specified distance from each other, a special holder was made out of Styrofoam was fabricated on CNC milling machine. The dielectric permittivity of Styrofoam is below 1.1, so its effective influence on the electromagnetic field distribution in a chain can be neglected. The thickness of the holder is chosen to be enough to ensure that the near-field radiation of the sample does not interact with the table on which the sample is placed.

\bibliography{apssamp}

\end{document}